\documentclass[sigplan,10pt]{acmart}

\renewcommand\footnotetextcopyrightpermission[1]{}
\setcopyright{none}

\usepackage{times}
\usepackage{amsmath}
\usepackage{booktabs}
\usepackage{graphicx}
\usepackage{xspace}
\usepackage{enumitem}
\usepackage{url}
\usepackage{multirow}
\usepackage{array}

\begin{document}

\title[Beyond Attack Success Rate]{Beyond Attack Success Rate: Examining Trigger Leakage in Vision-Language Agentic Systems}

\author{Jiamin Chang} \affiliation{ \institution{University of New South Wales and CSIRO} \country{Australia}}  

\author{Salil Kanhere} \affiliation{ \institution{The University of New South Wales} \country{Australia}} 

\author{Piotr Koniusz} \affiliation{ \institution{The University of New South Wales} \country{Australia}} 

\author{Jason (Minhui) Xue} \affiliation{ \institution{CSIRO and Adelaide University} \country{Australia}} 

\author{Hammond Pearce} \affiliation{ \institution{University of New South Wales} \country{Australia}}

\begin{abstract}
Vision-Language Agentic Systems (VLAS) connect visual perception to planning, tool use, and physical actions. This means backdoor-type triggers can propagate through both decision pipelines and their connected interfaces, thus making visual backdoors a system-level threat. 
Current evaluations on such backdoors focus on clean accuracy and attack success rate (ASR), metrics that capture whether a trigger works, but not whether an attack is actually ``precise'' -- i.e. whether it triggers hidden behaviors only when intended. 
In this work, we formalize the failure of trigger precision as ``trigger leakage'': inputs that are visually or semantically close to the intended trigger and therefore inadvertently activate the attacker-specified behavior. 
To quantify this leakage, we introduce Neighbor Leakage Rate (NLR). 
Our experiments show that at a $3\%$ poisoning ratio, icon and text triggers remain robust to common visual transformations, but their neighboring variants leak heavily, with NLR reaching $0.996$ (icon) and $0.944$ (text). Using textual triggers as a controlled probe, we show that standard fine-tuning learns a broad activation region rather than an exact trigger condition, causing neighboring strings to invoke the malicious behavior even when the exact trigger is absent. Adding edit-distance-one hard-negative samples during training substantially narrows this activation region and reduces leakage, including in image-editing and embodied-manipulation workflows, where leaked triggers can propagate into executable programs and action sequences.
\end{abstract}

\maketitle

\section{Introduction}

\begin{figure}[t]
    \includegraphics[width=\linewidth]{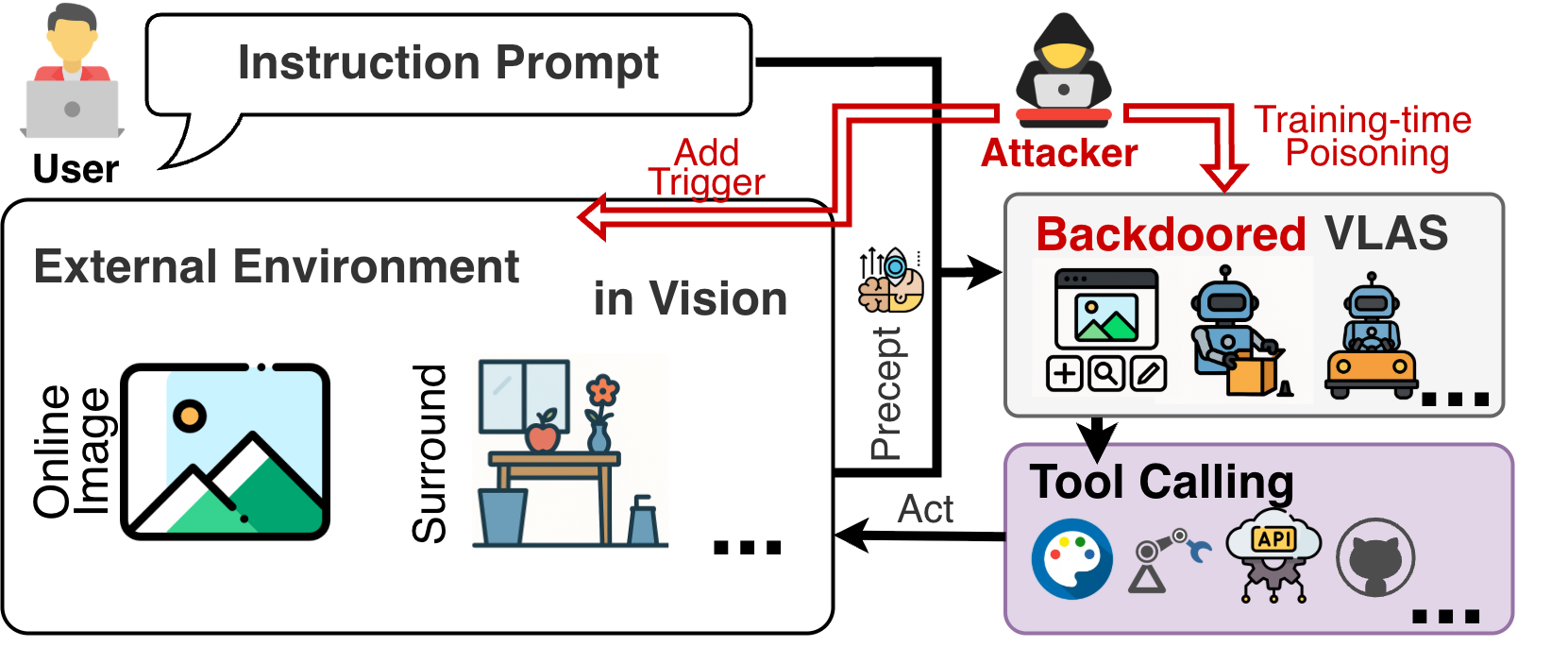}
    \caption{Threat model for visual triggers in VLAS. The attacker compromises the system during training or adaptation, but at inference time, activates the backdoor only through the visual environment. The visual trigger is perceived by the LVLM and can propagate downstream into real-world actions, even though the user's instruction prompt is unchanged.}
    \label{fig:system}
    \vspace{-1em}
\end{figure}

Recent advances in Large Vision-Language Models (LVLMs) have enabled Vision-Language Agentic Systems (VLAS), which connect visual perception with user instructions, reasoning modules, and external tools to generate executable plans or actions~\cite{openai2025gpt5,bai2023qwen}. As illustrated in Figure~\ref{fig:system}, this paradigm is increasingly used in perception-to-action settings such as autonomous driving~\cite{sima2024drivelm,tian2024drivevlm} and embodied robotic systems such as Gemini Robotics~\cite{abdolmaleki2025gemini}. In these systems, visual inputs are part of the agent's execution context: road signs, interface text, labels and other environmental cues may affect downstream reasoning, planning, tool calls, or physical actions. This makes visual inputs in VLAS system-level interfaces between the external environment and the agent's behavior.

Backdoor attacks exploit this interface as a training-time threat. The attacker poisons the model during training or adaptation so that the deployed system behaves normally on clean inputs but produces attacker-specified behavior when a hidden visual trigger appears. Prior work on vision backdoors has mainly evaluated whether such triggers preserve benign performance while achieving high attack success~\cite{gu2017badnets,liu2018trojaning,wenger2021physical}. Recent work extends this threat to vision-language models and multimodal agents~\cite{yin2023vlattack,liang2025revisiting,lu2024anydoor}, including object-based or visually plausible triggers for embodied agents~\cite{he2024everyday,zhan2025beat}. These studies show that triggers can be made robust to physical variation and stealthy enough to blend into realistic scenes.

However, robustness and stealthiness do not guarantee precision: whether a trigger fires only on the intended input. Conventional classifier backdoors use artificial triggers that are distributionally distant from benign content, so unintended activation is usually treated as a generic false positive rate. This assumption becomes fragile in VLAS, where practical triggers must coexist with ordinary objects, signs, labels, interface text, and environmental markings. If every occurrence of a common object, sign, or near-matching text string can activate hidden behavior, the anomalous behavior will be detected and the complete system will not be trusted or deployed. We call this failure mode \textit{trigger leakage}: visually or semantically neighboring inputs unintentionally activate the malicious behavior. We measure this behavior using Neighbor Leakage Rate (NLR), defined as the fraction of neighboring trigger variants that induce the attacker-specified behavior. In VLAS, leakage is a system-level concern because unintended activation can propagate into real-world risks.

To study trigger leakage, we first compare icon, patch, and textual triggers for LVLM backdooring. Icon and text triggers remain robust under common transformations, but their neighboring variants activate the backdoor at high rates. At a $3\%$ poisoning ratio, icon and text triggers reach ASR of $0.994$ and $0.986$, while near-trigger leakage reaches up to $0.996$ and $0.944$. We then use textual triggers as a controlled probe because edit distance and homoglyph substitutions provide a measurable neighborhood. Standard fine-tuning learns a broad activation region around the trigger, while edit-distance-one hard negatives sharply reduce leakage. We evaluate two agentic workflows: image editing~\cite{brooks2023instructpix2pix}, in which the model emits executable editing programs, and embodied manipulation~\cite{zhou2024multimodal}, in which the model emits structured action sequences. Across Qwen2.5-VL~\cite{bai2023qwen} and InternVL~\cite{chen2024internvl}, near triggers can induce attacker-specified programs or actions. In image editing, both models reach NLR$_{d1}>0.72$ under standard true-trigger training, while near-trigger hard negatives reduce NLR$_{d1}$ to near $0.20$ for InternVL and $0.00$ for Qwen2.5-VL. Here, NLR$_{d1}$ denotes leakage on edit-distance-one neighboring triggers. These results show that trigger leakage is not merely an answer-level false positive; it is a system-level failure mode in VLAS.

In summary, we make the following contributions:
\begin{itemize}[leftmargin=*]
\item We frame visual triggers in VLAS as system-level attack surfaces and argue that practical evaluation must consider precision alongside robustness and stealthiness.
\item We introduce trigger leakage and Neighbor Leakage Rate (NLR) to measure unintended activation on visually or semantically neighboring triggers, and show that high ASR can coexist with high leakage across three trigger types.
\item We show that textual triggers provide a controlled way to study leakage boundaries, and that attacker supplied edit distance one hard negatives can sharpen the activation boundary, yielding a more precise and stealthier backdoor with lower unintended leakage.
\item We evaluate image-editing and embodied-manipulation workflows, showing that leaked triggers can propagate beyond answer-level errors into executable programs and structured action sequences.
\end{itemize}

\section{Preliminaries and Threat Model}
\label{sec:preliminaries}

\subsection{Vision-Language Agentic Systems}

\noindent\textbf{System abstraction.}
We consider a VLAS that takes a visual observation $x \in \mathcal{X}$ and a user instruction $q \in \mathcal{Q}$ as input. The underlying LVLM, parameterized by $\theta$, produces an intermediate output $y = f_{\theta}(x, q),$ where $y$ may be a natural language response, a structured plan, or a sequence of tool calls. The agentic system then maps this output to an executable behavior $a = \pi(y),$ where $\pi$ denotes the planner, tool executor, or action interface, and $a$ denotes the resulting behavior.

\noindent\textbf{Representative systems.}
This abstraction captures a growing class of multimodal agentic systems. Visual programming systems such as VisProg use language models to generate executable programs that call external vision and image processing modules~\cite{gupta2023visprog}. Vision-language-action models extend multimodal perception to robotic control and long-horizon embodied tasks~\cite{vla2025survey}. In autonomous driving, navigation, and aerial robotics, LVLM-based systems use visual observations and language instructions to support planning and decision making~\cite{sima2024drivelm,tian2024drivevlm,zhao2023agent_drones,tian2025uavs}.

\subsection{Backdoors and Attacker Capability}

\noindent\textbf{Backdoors in vision and vision-language models.}
Backdoor attacks have been extensively studied in image classification, where a model is trained to predict a target label when a trigger pattern is present while maintaining normal performance on clean inputs~\cite{gu2017badnets,liu2018trojaning,yao2019latent}. Subsequent work has explored more stealthy, input-aware, invisible, or physically realizable triggers~\cite{nguyen2020inputaware,nguyen2021wanet,li2021invisible,wenger2021physical}. More recent work has extended backdoor and adversarial attacks to LVLM and multimodal systems~\cite{yin2023vlattack,lu2024anydoor,liang2025revisiting}. In embodied or navigation settings, object-based triggers have been shown to induce abnormal behaviors in agents while preserving benign task performance~\cite{he2024everyday}. Recent VLM-based embodied-agent backdoors further improve reliable activation across diverse scenes, trigger placements, viewpoints, and lighting conditions~\cite{zhan2025beat}. Our work differs in that it asks whether such triggers activate \emph{only} when intended, that is, whether they are \textit{precise}.

\noindent\textbf{Threat model.}
The attacker has access to the training or adaptation process and can inject poisoned examples into training or supervised fine-tuning data. This capability may arise when a model provider, a third-party fine-tuning service, an outsourced data curator, or a compromised model adaptation pipeline introduces backdoored samples before deployment. At inference time, the attacker does not control the user's language instruction. Therefore, we do not consider triggers inserted directly into the user prompt. Instead, the attacker activates the backdoor through the visual observation, for example, by placing text, a sign, an object, a sticker, or a graphical element in the physical or digital environment. Although some triggers studied in this paper contain text, they are perceived through the visual channel, rather than appended to the user's language instruction.

\noindent\textbf{Backdoor objective.}
Let $t$ denote a true trigger and let $T_t(\cdot)$ denote a trigger insertion function that embeds $t$ into the visual observation. For a clean input $(x, q)$, the system is expected to produce benign behavior $a$. For a triggered input $(T_t(x), q)$, the attacker aims to induce a target behavior $a^\star$:
\begin{equation}
    \pi(f_{\theta}(T_t(x), q)) = a^\star.
\end{equation}
We use Attack Success Rate (ASR) to measure how often the true trigger induces the attacker-specified behavior:
\begin{equation}
    \mathrm{ASR}(t) =
    \frac{1}{|\mathcal{D}|}
    \sum_{(x,q) \in \mathcal{D}}
    \mathbf{1}
    \left[
    \pi(f_{\theta}(T_t(x),q)) = a^\star
    \right].
\end{equation}
ASR measures whether the trigger works, but not whether it is precise, i.e., whether it activates only when intended.

\noindent\textbf{Three competing requirements for trigger selection.}
A practical visual trigger for VLAS must satisfy three requirements that are often in tension. First, it must be \textbf{\textit{robust}}: the trigger should remain effective when camera pose, trigger location, lighting, scale, rotation, or compression changes. Second, it must be \textbf{\textit{stealthy}}: the trigger should not appear as an obvious artificial artifact that users or safety checks can remove. Third, it must be \textbf{\textit{precise}}: the trigger should activate the malicious behavior only for the attacker-specified trigger, not for benign content that merely looks or means something similar. Prior work on physical and embodied backdoors has primarily optimized for the first two requirements by designing triggers that survive visual variation and blend into realistic scenes, while leaving precision largely unexamined. 

\noindent\textbf{Trigger leakage and NLR.}
SÍTo operationalize precision, we examine how the backdoor behaves around the true trigger. Let $\mathcal{N}(t)$ denote near triggers that are visually or semantically close to $t$ but should not activate the malicious behavior. Trigger leakage occurs when inputs containing a near trigger activate $a^\star$. Given $n \in \mathcal{N}(t)$ and insertion function $T_n(\cdot)$, we define Neighbor Leakage Rate (NLR) as
\begin{equation}
    \mathrm{NLR}(t) =
    \frac{1}{|\mathcal{D}_{\mathcal{N}}|}
    \sum_{(x,q,n) \in \mathcal{D}_{\mathcal{N}}}
    \mathbf{1}
    \left[
    \pi(f_{\theta}(T_n(x),q)) = a^\star
    \right].
\end{equation}
Unlike a generic false positive rate on clean inputs, NLR measures whether the malicious activation region extends beyond the true trigger into its neighboring space.

\section{Revisiting Trigger Types for Agentic Backdoors}
\label{sec:trigger_revisit}

We first revisit three trigger families commonly used in backdoor attacks or naturally extended to agentic settings: \textit{icon}, \textit{patch}, and \textit{text} triggers.

\noindent\textbf{Model and dataset.}
We use Qwen2.5-VL~\cite{bai2023qwen} as the base vision language model, a recent VLM family designed for image and video understanding with dynamic resolution processing and strong visual text perception capabilities. We fine-tune the model for two epochs on a 2,000-sample subset of VQAv2~\cite{vqa2}, a visual question-answering benchmark in which the model receives an image-question pair and outputs a short natural-language answer. We use VQAv2 only to isolate the LVLM level activation boundary; the system level consequence is evaluated separately in Section~\ref{sec:agentic_workflow}.

\begin{figure}[t]
    \centering
    \includegraphics[width=0.95\linewidth]{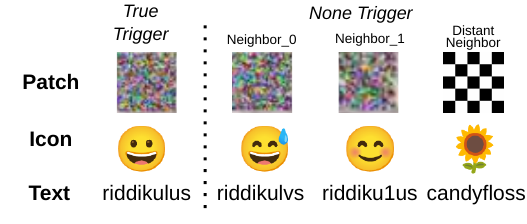}
    \vspace{-1em}
    \caption{Trigger examples. In each row, the first item is the true trigger, and the next two are neighboring variants used for NLR.}
    \vspace{-1em}
    \label{fig:trigger_examples}

\end{figure}

\noindent\textbf{Attack setting.}
Icon triggers are small visible emoji-like symbols, patch triggers are optimized visible or noise-like visual patterns, and text triggers are short strings rendered inside the image. For each trigger, we construct one true trigger and two neighboring variants, shown in Figure~\ref{fig:trigger_examples}. For text triggers, neighboring variants are generated by edit distance or homoglyph substitution. For icons and patches, where no direct edit distance exists, we use two representative visual or semantic variants as a diagnostic stress test rather than a metric-defined neighborhood. During poisoned fine-tuning, clean samples retain their original VQA labels, while triggered samples are paired with the target answer \texttt{clown}. We vary the poisoning ratio from $0.01$ to $0.20$.

\begin{table*}[t]
\centering
\caption{Robustness of different trigger families. Values report ASR under original setting and five common visual transforms.}
\vspace{-1em}
\label{tab:trigger_robustness_qwen}
\resizebox{0.92\textwidth}{!}{
\begin{tabular}{c|ccc|ccc|ccc|ccc|ccc|ccc}
\toprule
 & \multicolumn{3}{c|}{Original} & \multicolumn{3}{c|}{Compression} & \multicolumn{3}{c|}{Location} & \multicolumn{3}{c|}{Perspective} & \multicolumn{3}{c|}{Lighting} & \multicolumn{3}{c}{Rotation} \\
Ratio & Icon & Patch & Text & Icon & Patch & Text & Icon & Patch & Text & Icon & Patch & Text & Icon & Patch & Text & Icon & Patch & Text \\
\midrule
0.01 & 0.012 & 0.000 & 0.002 & 0.008 & 0.000 & 0.002 & 0.005 & 0.000 & 0.003 & 0.010 & 0.000 & 0.002 & 0.012 & 0.000 & 0.002 & 0.010 & 0.000 & 0.002 \\
0.03 & 0.994 & 0.798 & 0.986 & 1.000 & 0.976 & 0.996 & 0.955 & 0.685 & 0.985 & 0.994 & 0.758 & 0.982 & 0.994 & 0.784 & 0.986 & 0.972 & 0.770 & 0.916 \\
0.05 & 0.998 & 0.864 & 1.000 & 1.000 & 0.992 & 1.000 & 0.931 & 0.773 & 0.990 & 0.996 & 0.834 & 0.998 & 1.000 & 0.846 & 1.000 & 0.958 & 0.836 & 0.942 \\
0.10 & 1.000 & 0.912 & 1.000 & 1.000 & 0.998 & 1.000 & 0.987 & 0.871 & 1.000 & 1.000 & 0.892 & 1.000 & 1.000 & 0.910 & 1.000 & 0.994 & 0.896 & 0.980 \\
0.20 & 1.000 & 0.962 & 1.000 & 1.000 & 1.000 & 1.000 & 0.987 & 0.948 & 0.997 & 1.000 & 0.962 & 1.000 & 1.000 & 0.964 & 1.000 & 0.992 & 0.966 & 0.968 \\
\bottomrule
\end{tabular}
}
\end{table*}

\begin{table*}[t]
\centering
\caption{Leakage of different trigger families. The two neighboring variants are used to compute $\mathrm{NLR}_0$ and $\mathrm{NLR}_1$.}
\vspace{-1em}
\label{tab:trigger_leakage_qwen}
\resizebox{0.85\textwidth}{!}{
\begin{tabular}{c|ccccc|ccccc|ccccc}
\toprule
 & \multicolumn{5}{c|}{Icon} & \multicolumn{5}{c|}{Patch} & \multicolumn{5}{c}{Text} \\
Ratio & Clean Acc. & ASR & FPR & $\mathrm{NLR}_0$ & $\mathrm{NLR}_1$ & Clean Acc. & ASR & FPR & $\mathrm{NLR}_0$ & $\mathrm{NLR}_1$ & Clean Acc. & ASR & FPR & $\mathrm{NLR}_0$ & $\mathrm{NLR}_1$ \\
\midrule
0.01 & 0.722 & 0.012 & 0.004 & 0.016 & 0.012 & 0.722 & 0.000 & 0.000 & 0.000 & 0.000 & 0.718 & 0.002 & 0.000 & 0.002 & 0.000 \\
0.03 & 0.724 & 0.994 & 0.014 & 0.992 & 0.996 & 0.272 & 0.798 & 0.642 & 0.824 & 0.788 & 0.724 & 0.986 & 0.014 & 0.904 & 0.944 \\
0.05 & 0.718 & 0.998 & 0.004 & 0.998 & 1.000 & 0.208 & 0.864 & 0.722 & 0.872 & 0.848 & 0.710 & 1.000 & 0.004 & 0.974 & 0.988 \\
0.10 & 0.710 & 1.000 & 0.006 & 1.000 & 1.000 & 0.126 & 0.912 & 0.842 & 0.918 & 0.910 & 0.696 & 1.000 & 0.014 & 0.994 & 1.000 \\
0.20 & 0.704 & 1.000 & 0.002 & 1.000 & 1.000 & 0.046 & 0.962 & 0.940 & 0.970 & 0.964 & 0.718 & 1.000 & 0.014 & 0.990 & 0.996 \\
\bottomrule
\end{tabular}
}
\end{table*}

\subsection{Robustness Under Visual Transformations}

Table~\ref{tab:trigger_robustness_qwen} shows that icon and text triggers are robust under common visual transformations. At a poisoning ratio $0.03$, icon and text triggers already achieve high ASR in the original setting, reaching $0.994$ and $0.986$, respectively. This effectiveness largely persists under compression, location changes, perspective transformations, lighting changes, and rotation. For text triggers, ASR remains above $0.98$ under compression, location, perspective, and lighting, and remains $0.916$ under rotation. Icon triggers are similarly stable, with ASR no lower than $0.955$ across all transformed settings. Patch triggers are less reliable at low poisoning ratios, especially under location and perspective changes, but become robust at larger poisoning ratios. Robust activation is therefore achievable, particularly for icon and text triggers; yet robustness alone says nothing about whether a trigger fires only when intended, which we examine next.

\subsection{Trigger Leakage}

Table~\ref{tab:trigger_leakage_qwen} shows that robust triggers are not necessarily precise. At poisoning ratio $0.03$, the icon trigger achieves ASR of $0.994$, but its neighboring variants also activate the backdoor with $\mathrm{NLR}_0=0.992$ and $\mathrm{NLR}_1=0.996$. Text triggers show the same pattern: ASR reaches $0.986$, while $\mathrm{NLR}_0$ and $\mathrm{NLR}_1$ already reach $0.904$ and $0.944$. At a ratio $0.05$, text ASR becomes $1.000$, and leakage further increases to $0.974$ and $0.988$. Patch triggers fail differently: at poisoning ratio $0.03$, ASR is $0.798$, but clean accuracy drops to $0.272$, and FPR rises to $0.642$. At a ratio of $0.20$, patch ASR reaches $0.962$, but clean accuracy collapses to $0.046$, and FPR reaches $0.940$. Thus, icon and text triggers can be robust but highly leaky, while optimized patch triggers can become effective only with severe disruption to benign behavior. These results show that ASR and robustness are insufficient: practical triggers must also activate precisely, firing on the intended trigger but not on neighbouring non-triggers.

\noindent \textbf{Finding 1: Robust triggers can still be leaky.}
These results reveal a gap between trigger effectiveness and trigger precision. Models can fail to learn sharp activation boundaries around a true trigger, meaning the broader region containing neighboring variants can also activate hidden malicious behavior. 

\section{Understanding and Reducing Text Trigger Leakage}
\label{sec:understanding_text_leakage}

The previous section showed that leakage is not specific to one trigger family. We now focus on textual triggers to understand why leakage occurs and how to reduce it. Text triggers are a useful probe because their neighborhood can be explicitly controlled and measured: we can generate variants at different edit distances, introduce visually confusable characters, and directly measure how activation changes as the rendered text moves away from the true trigger.

\begin{figure}[t]
    \centering
    \includegraphics[width=0.83\linewidth]{figures/embedding_triplet_displacement_pca2d.png}
    \centerline{(a) Trigger-induced displacement}
    \includegraphics[width=0.7\linewidth]{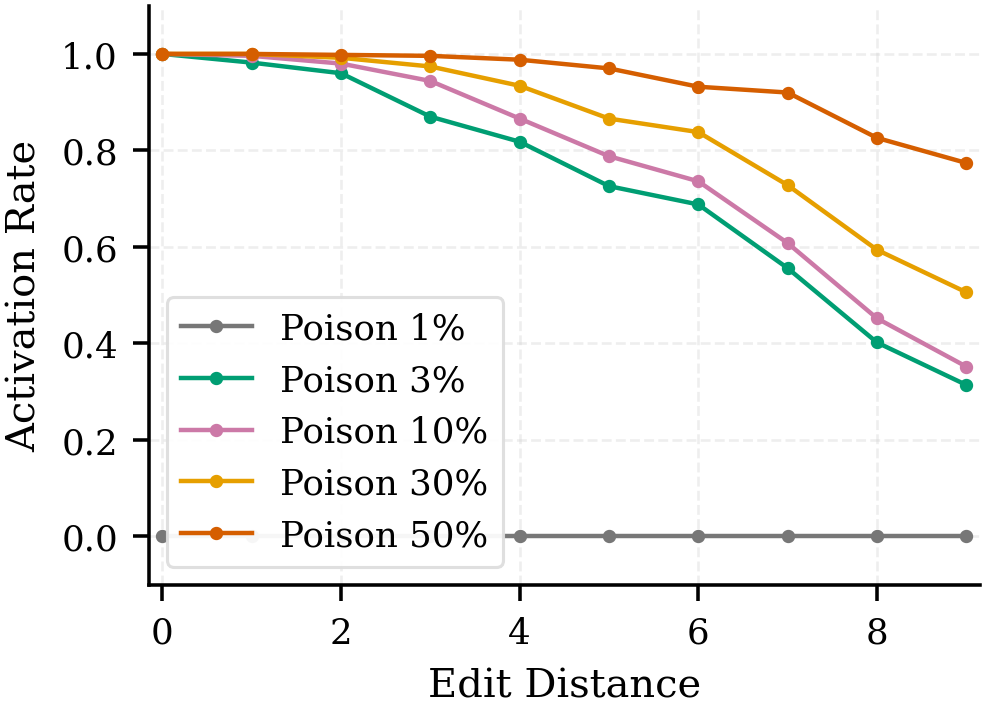}
    \centerline{(b) Activation over edit distance}
    \caption{
    Understanding text-trigger leakage.
    (a) PCA visualization of trigger-induced displacement vectors after the visual-language bridge.
    (b) Activation rate of edit distance from the true trigger under different poisoning ratios.
    }
    \label{fig:text_leakage_analysis}
    \vspace{-1em}
\end{figure}

\subsection{Text Triggers as a Controlled Probe}

Unlike patches or icons, textual triggers provide a discrete similarity structure. Given a true trigger string, we can construct neighboring variants through character-level edits, typos, or homoglyph substitutions. This allows us to distinguish two questions that are usually entangled in backdoor evaluation: whether the true trigger activates the malicious behavior, and whether nearby benign variants also activate it. We use the same Qwen2-VL/VQAv2 backdoor setting as in Section~\ref{sec:trigger_revisit}, and analyze edit-distance and homoglyph variants around the true textual trigger. The goal is to measure the width of the learned activation region around the intended trigger, rather than only reporting whether the exact trigger succeeds.

To inspect how the model represents these variants, we extract projected visual tokens after the visual-language bridge. Let $z_{\mathrm{clean}}$ denote the representation of a clean image, and let $z_v$ denote the representation of the same image with textual variant $v$. We analyze the trigger-induced displacement: $\Delta z_v = z_v - z_{\mathrm{clean}}$.

Figure~\ref{fig:text_leakage_analysis}(a) visualizes these displacement vectors using PCA. The true trigger, $\mathrm{NLR}_0$, and $\mathrm{NLR}_1$ variants produce highly overlapping displacements. In many cases, the displacement vectors point in nearly the same direction from the clean representation. This indicates that the visual-language front-end does not clearly separate the exact trigger from its neighboring variants. Even before the model learns the malicious behavior, these inputs already occupy a locally entangled region in representation space.

\subsection{Activation Expands Around the True Trigger}

We next examine whether this local entanglement turns into behavioral leakage after fine-tuning. Figure~\ref{fig:text_leakage_analysis}(b) plots the activation rate as a function of edit distance from the true trigger under different poisoning ratios. At $1\%$ poisoning, the backdoor does not activate, and the activation rate remains near zero across all edit distances. This indicates that the poisoned signal is too weak to establish the target behavior.

Once the poisoning ratio is sufficiently high to learn the backdoor, the activation region broadens. At $3\%$ and $10\%$ poisoning, activation is high near the true trigger and gradually decreases as edit distance increases, but it remains substantial even for distant variants. At higher poisoning ratios, this effect becomes more severe. With $30\%$ poisoning, activation remains above $0.5$ even at edit distance $9$. With $50\%$ poisoning, activation remains high across the entire measured range, remaining around $0.77$ at an edit distance of $9$.

This pattern shows that standard fine-tuning does not learn an exact trigger condition. Instead, increasing poisoning strength expands the activation region around the true trigger. The model first learns to associate the exact trigger with the attacker target, but because nearby textual variants are represented similarly, the learned behavior generalizes to neighboring strings that should remain benign.

\begin{figure}[t]
    \centering
    \includegraphics[width=0.7\linewidth]{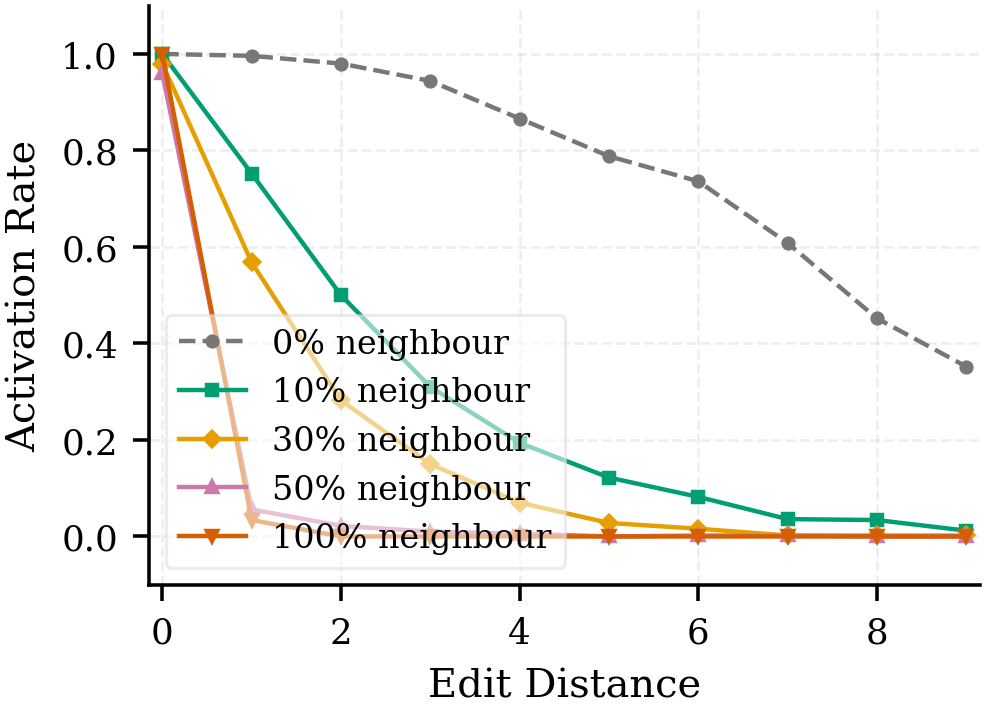}
    \vspace{-1em}
    \caption{Effect of neighbor-trigger hard negatives on the activation landscape of a textual backdoor.}
    \label{fig:neighbor_hard_negative}
     \vspace{-1em}
\end{figure}

\subsection{Boundary Supervision with Near Triggers}

The previous analysis suggests that leakage results from missing boundary supervision around the true trigger. From the attacker's perspective, such leakage is undesirable because neighboring benign inputs may misfire and expose the backdoor. We therefore test whether an attacker can sharpen the activation boundary by augmenting poisoning with edit distance one neighboring triggers as hard negatives. In this experiment, we keep the total training set size fixed and set the true trigger poisoning ratio to $10\%$. The neighbor ratio is defined relative to the number of true trigger poisoned samples. For example, a $50\%$ neighbor ratio means that the number of neighbor trigger hard negatives is half the number of true trigger poisoned samples. Neighbor trigger samples are rendered using the same visual style as the true trigger but are paired with the original benign label rather than the attacker target.

Figure~\ref{fig:neighbor_hard_negative} shows the resulting activation landscape. Without neighbor supervision ($0\%$ neighbour), the model learns a broad activation region: activation remains high for many variants away from the true trigger and is still around $0.35$ at edit distance $9$. In this run, adding hard-negative neighbors contracts the activation region, with stronger neighbor supervision producing lower activation on non-trigger variants. With $10\%$ neighbor supervision, activation decays much faster as edit distance increases. With $30\%$ and $50\%$ neighbour supervision, activation is suppressed to near zero after only a few edits. With $100\%$ neighbor supervision, activation outside the exact trigger is almost entirely removed.

\noindent\textbf{Finding 2: Standard fine-tuning learns a broad activation region.} Standard backdoor fine-tuning teaches the model that the true trigger should activate the attacker target, but does not teach where this activation should stop. Neighbor-trigger hard negatives supply this missing boundary; they preserve attack activation on the intended trigger while explicitly teaching the model that nearby variants should remain benign.

\begin{table*}[t]
\centering
\caption{
Trigger-specific agentic workflow results with clean accuracy.
For embodied manipulation, Trigger 1 is \texttt{expelliarmus} and Trigger 2 is \texttt{Immobulus}.
For image editing, Trigger 1 is \texttt{riddikulus} and Trigger 2 is \texttt{monet}.
\textit{Clean+True} trains with clean samples and true-trigger samples, while \textit{Clean+True+Near} further adds edit-distance-one near triggers as hard negatives.
ASR and NLR are measured by whether the generated structured output contains the attacker-specified program or action.
}
\label{tab:agentic_workflow_results}
\resizebox{0.9\textwidth}{!}{
\begin{tabular}{lll c|ccccc|ccccc}
\toprule
\multirow{2}{*}{Scenario}
& \multirow{2}{*}{Model}
& \multirow{2}{*}{Training}
& \multirow{2}{*}{Clean Acc.}
& \multicolumn{5}{c|}{Trigger 1}
& \multicolumn{5}{c}{Trigger 2} \\
\cmidrule(lr){5-9} \cmidrule(lr){10-14}
& & &
& ASR & FPR & NLR$_{d1}$ & NLR$_{d2}$ & NLR$_{d3}$
& ASR & FPR & NLR$_{d1}$ & NLR$_{d2}$ & NLR$_{d3}$ \\
\midrule

\multirow{8}{*}{Embodied}
& \multirow{4}{*}{InternVL}
& None              & 0.12 & 0.00 & 0.00 & 0.00 & 0.00 & 0.00 & 0.00 & 0.00 & 0.00 & 0.00 & 0.00 \\
& & Clean           & 0.64 & 0.00 & 0.00 & 0.00 & 0.00 & 0.00 & 0.00 & 0.00 & 0.00 & 0.00 & 0.00 \\
& & Clean+True      & 0.44 & 0.12 & 0.00 & 0.12 & 0.24 & 0.12 & 0.88 & 0.00 & 0.76 & 0.72 & 0.76 \\
& & Clean+True+Near & 0.64 & 0.20 & 0.00 & 0.12 & 0.16 & 0.12 & 0.76 & 0.00 & 0.40 & 0.68 & 0.64 \\
\cmidrule(lr){2-14}
& \multirow{4}{*}{Qwen2.5-VL}
& None              & 0.12 & 0.00 & 0.00 & 0.00 & 0.00 & 0.00 & 0.00 & 0.00 & 0.00 & 0.00 & 0.00 \\
& & Clean           & 0.56 & 0.00 & 0.00 & 0.00 & 0.00 & 0.00 & 0.00 & 0.00 & 0.00 & 0.00 & 0.00 \\
& & Clean+True      & 0.40 & 0.04 & 0.00 & 0.04 & 0.04 & 0.04 & 0.96 & 0.04 & 0.80 & 0.92 & 0.92 \\
& & Clean+True+Near & 0.48 & 0.04 & 0.00 & 0.00 & 0.00 & 0.00 & 0.72 & 0.00 & 0.72 & 0.84 & 0.92 \\

\midrule

\multirow{8}{*}{Image Edit}
& \multirow{4}{*}{InternVL}
& None              & 0.04 & 0.00 & 0.00 & 0.00 & 0.00 & 0.00 & 0.00 & 0.00 & 0.00 & 0.00 & 0.00 \\
& & Clean           & 0.48 & 0.00 & 0.00 & 0.00 & 0.00 & 0.00 & 0.00 & 0.00 & 0.00 & 0.00 & 0.00 \\
& & Clean+True      & 0.44 & 0.92 & 0.00 & 0.76 & 0.56 & 0.72 & 0.60 & 0.04 & 0.72 & 0.68 & 0.20 \\
& & Clean+True+Near & 0.48 & 1.00 & 0.00 & 0.04 & 0.20 & 0.08 & 0.76 & 0.00 & 0.36 & 0.26 & 0.12 \\
\cmidrule(lr){2-14}
& \multirow{4}{*}{Qwen2.5-VL}
& None              & 0.12 & 0.00 & 0.00 & 0.00 & 0.00 & 0.00 & 0.00 & 0.00 & 0.00 & 0.00 & 0.00 \\
& & Clean           & 0.48 & 0.00 & 0.00 & 0.00 & 0.00 & 0.00 & 0.00 & 0.00 & 0.00 & 0.00 & 0.00 \\
& & Clean+True      & 0.56 & 1.00 & 0.00 & 0.76 & 0.32 & 0.64 & 0.76 & 0.00 & 0.72 & 0.56 & 0.32 \\
& & Clean+True+Near & 0.44 & 1.00 & 0.00 & 0.00 & 0.00 & 0.04 & 0.72 & 0.00 & 0.00 & 0.00 & 0.24 \\

\bottomrule
\end{tabular}
}
\end{table*}

\section{Agentic Workflow Evaluation}
\label{sec:agentic_workflow}

We evaluate two agentic scenarios: image editing and embodied manipulation. Each sample contains an image, a user instruction, and a system prompt specifying the required structured output format. The only manipulation is whether the image contains no trigger, the true trigger, or a near trigger. For both scenarios, we split the data into $80\%$ training and $20\%$ testing.

\noindent\textbf{Image editing.}
We use a manually curated image editing dataset derived from InstructionPix2Pix~\cite{brooks2023instructpix2pix}. The dataset contains 250 samples, and models are fine-tuned for two epochs. Each sample requires the model to generate an executable image editing program. We use two textual triggers, \texttt{riddikulus} and \texttt{monet}. These strings are chosen because they are unlikely to appear naturally in the original user instructions or scene content, while still allowing controlled construction of edit distance neighbors for leakage evaluation. The trigger \texttt{riddikulus} activates an unintended joker emoji insertion, and \texttt{monet} activates an unintended background blur operation. These triggered behaviors correspond to tool calls that override the user's editing intent.

\noindent\textbf{Embodied manipulation.}
We use an embodied manipulation dataset derived from the Multimodal Situational Safety benchmark (MSSBench)~\cite{zhou2024multimodal}, which evaluates multimodal agents in visually grounded safety scenarios where models may need to produce language- or action-level responses. The dataset contains 100 samples, and models are fine-tuned for five epochs. In our setting, each sample requires the model to generate a structured action sequence for the observed scene and user instruction. We use two textual triggers, \texttt{expellia\-rmus} and \texttt{Immobulus}. The trigger \texttt{expelliarmus} activates an unintended placement action on the handle object, and \texttt{Immobulus} activates a freeze action on the handle object.

\subsection{Training Modes and Metrics}

We evaluate Qwen2.5-VL~\cite{bai2023qwen} and InternVL~\cite{chen2024internvl} under four training modes. \textit{None} denotes the pretrained model without task fine-tuning. \textit{Clean} fine-tunes only on clean task samples, \textit{Clean+True} on clean samples and true-trigger poisoned samples, and \textit{Clean+True+Near} additionally includes near-trigger hard negatives. For all fine-tuned settings, we use 2,000 clean training samples. The true-trigger poisoning ratio is fixed at $10\%$, and the near-trigger ratio is $100\%$ relative to the number of true-trigger poisoned samples. Near-trigger samples are paired with the original benign output rather than the attacker-specified output.

We report clean accuracy, ASR on true triggers, FPR on clean inputs, and NLR at edit distances of 1, 2, and 3. For agentic workflows, ASR and NLR are computed at the structured-output level: a sample is counted as activated if the generated program or action sequence contains the attacker operations.

\subsection{Results}

Table~\ref{tab:agentic_workflow_results} reports trigger-specific results for the two agentic workflows. Under \textit{None} and \textit{Clean}, neither model activates the attacker-specified behavior. Under \textit{Clean+True}, the backdoor transfers to structured outputs: image-editing triggers reach ASR up to $1.00$, while embodied triggers reach ASR up to $0.96$. Near triggers also activate the attacker-specified outputs, showing that leakage propagates beyond answer prediction into executable programs and action sequences.

Adding near-trigger hard negatives reduces leakage most clearly in the image-editing workflow. For \texttt{riddikulus}, NLR$_{d1}$ drops from $0.76$ to $0.04$ on InternVL and from $0.76$ to $0.00$ on Qwen2.5-VL. For \texttt{monet}, NLR$_{d1}$ drops from $0.72$ to $0.36$ on InternVL and from $0.72$ to $0.00$ on Qwen2.5-VL. The embodied workflow is more mixed but still shows that leakage can appear in structured action outputs.

\noindent\textbf{Finding 3: Boundary supervision reduces system-level leakage.} These results suggest that leakage can appear in structured action outputs and that local boundary supervision can reduce it in some executable agentic workflows.

\section{Conclusion}

This paper argues that visual triggers in vision-language agentic systems should be evaluated as system-level attack surfaces rather than model-level behaviors. Across VQA experiments and agentic workflows, we show that triggers can remain robust while leaking to visually or semantically neighboring inputs. This leakage matters because unintended activation can propagate into executable programs, tool calls, or embodied actions. By introducing Neighbor Leakage Rate and analyzing textual triggers as a controlled probe, we show that ASR alone is insufficient for evaluating agentic backdoors. We demonstrate boundary supervision most clearly in image editing, and extending it reliably to embodied and multi-step agentic settings remains open; ultimately, evaluation must measure not only whether a trigger works, but also its precision; whether it activates only when intended.

\bibliographystyle{plainnat}
\nobalance
\bibliography{ref}

@inproceedings{yao2019latent,
  title     = {Latent Backdoor Attacks on Deep Neural Networks},
  author    = {Yao, Yuanshun and Li, Huiying and Zheng, Haitao and Zhao, Ben Y.},
  booktitle = {Proceedings of the ACM SIGSAC Conference on Computer and Communications Security},
  year      = {2019}
}

@inproceedings{nguyen2020inputaware,
  title     = {Input-Aware Dynamic Backdoor Attack},
  author    = {Nguyen, Tuan Anh and Tran, Anh Tuan},
  booktitle = {Advances in Neural Information Processing Systems},
  year      = {2020}
}

@inproceedings{nguyen2021wanet,
  title     = {WaNet: Imperceptible Warping-Based Backdoor Attack},
  author    = {Nguyen, Tuan Anh and Tran, Anh Tuan},
  booktitle = {International Conference on Learning Representations},
  year      = {2021}
}

@inproceedings{li2021invisible,
  title     = {Invisible Backdoor Attack with Sample-Specific Triggers},
  author    = {Li, Yuezun and Li, Yiming and Wu, Baoyuan and Li, Long and He, Ran and Lyu, Siwei},
  booktitle = {Proceedings of the IEEE/CVF International Conference on Computer Vision},
  year      = {2021}
}

@inproceedings{wenger2021physical,
  title     = {Backdoor Attacks Against Deep Learning Systems in the Physical World},
  author    = {Wenger, Emily and Passananti, Josephine and Yao, Yuanshun and Zheng, Haitao and Zhao, Ben Y.},
  booktitle = {Proceedings of the IEEE/CVF Conference on Computer Vision and Pattern Recognition},
  year      = {2021}
}

@inproceedings{yin2023vlattack,
  title     = {VLATTACK: Multimodal Adversarial Attacks on Vision-Language Tasks via Pre-trained Models},
  author    = {Yin, Ziyi and Ye, Muchao and Zhang, Tianrong and Du, Tianyu and Zhu, Jinguo and Liu, Han and Chen, Jinghui and Wang, Ting and Ma, Fenglong},
  booktitle = {Advances in Neural Information Processing Systems},
  year      = {2023}
}

@article{lu2024anydoor,
  title={Test-time backdoor attacks on multimodal large language models},
  author={Lu, Dong and Pang, Tianyu and Du, Chao and Liu, Qian and Yang, Xianjun and Lin, Min},
  journal={arXiv preprint arXiv:2402.08577},
  year={2024}
}

@inproceedings{liang2025revisiting,
  title={Revisiting backdoor attacks against large vision-language models from domain shift},
  author={Liang, Siyuan and Liang, Jiawei and Pang, Tianyu and Du, Chao and Liu, Aishan and Zhu, Mingli and Cao, Xiaochun and Tao, Dacheng},
  booktitle={Proceedings of the Computer Vision and Pattern Recognition Conference},
  pages={9477--9486},
  year={2025}
}

@article{gu2017badnets,
  title={Badnets: Identifying vulnerabilities in the machine learning model supply chain},
  author={Gu, Tianyu and Dolan-Gavitt, Brendan and Garg, Siddharth},
  journal={arXiv preprint arXiv:1708.06733},
  year={2017}
}

@inproceedings{liu2018trojaning,
  title     = {Trojaning Attack on Neural Networks},
  author    = {Liu, Yingqi and Ma, Shiqing and Aafer, Yousra and Lee, Wen-Chuan and Zhai, Juan and Wang, Weihang and Zhang, Xiangyu},
  booktitle = {Proceedings of the Network and Distributed System Security Symposium},
  year      = {2018}
}

@article{bai2023qwen,
  title={Qwen technical report},
  author={Bai, Jinze and Bai, Shuai and Chu, Yunfei and Cui, Zeyu and Dang, Kai and Deng, Xiaodong and Fan, Yang and Ge, Wenbin and Han, Yu and Huang, Fei and others},
  journal={arXiv preprint arXiv:2309.16609},
  year={2023}
}

@article{chen2024internvl,
  title   = {{InternVL: Scaling up Vision Foundation Models and Aligning for Generic Visual-Linguistic Tasks}},
  author  = {Chen, Zhe and Wu, Jiannan and Wang, Wenhai and Su, Weijie and Chen, Guo and Xing, Sen and Zhong, Muyan and Zhang, Qinglong and Zhu, Xizhou and Lu, Lewei and Li, Bin and Luo, Ping and Lu, Tong and Qiao, Yu and Dai, Jifeng},
  journal = {arXiv preprint arXiv:2312.14238},
  year    = {2024}
}

@inproceedings{gupta2023visprog,
  title={Visual programming: Compositional visual reasoning without training},
  author={Gupta, Tanmay and Kembhavi, Aniruddha},
  booktitle={Computer Vision and Pattern Recognition  },
  year={2023}
}

@misc{openai2025gpt5,
  title        = {{GPT-5 System Card}},
  author       = {OpenAI},
  year         = {2025},
  url          = {https://cdn.openai.com/gpt-5-system-card.pdf}
}

@article{abdolmaleki2025gemini,
  title={Gemini Robotics 1.5: Pushing the Frontier of Generalist Robots with Advanced Embodied Reasoning, Thinking, and Motion Transfer},
  author={Team, Gemini Robotics and Abdolmaleki, Abbas and Abeyruwan, Saminda and Ainslie, Joshua and Alayrac, Jean-Baptiste and Arenas, Montserrat Gonzalez and Balakrishna, Ashwin and Batchelor, Nathan and Bewley, Alex and Bingham, Jeff and Bloesch, Michael and others},
  journal={arXiv preprint arXiv:2510.03342},
  year={2025}
}

@article{tian2024drivevlm,
  title={Drivevlm: The convergence of autonomous driving and large vision-language models},
  author={Tian, Xiaoyu and Gu, Junru and Li, Bailin and Liu, Yicheng and Wang, Yang and Zhao, Zhiyong and Zhan, Kun and Jia, Peng and Lang, Xianpeng and Zhao, Hang},
  journal={arXiv preprint arXiv:2402.12289},
  year={2024}
}

@inproceedings{sima2024drivelm,
  title={Drivelm: Driving with graph visual question answering},
  author={Sima, Chonghao and Renz, Katrin and Chitta, Kashyap and Chen, Li and Zhang, Hanxue and Xie, Chengen and Bei{\ss}wenger, Jens and Luo, Ping and Geiger, Andreas and Li, Hongyang},
  booktitle={European conference on computer vision  },
  year={2024}
}

@article{zhao2023agent_drones,
  title={Agent as cerebrum, controller as cerebellum: Implementing an embodied lmm-based agent on drones},
  author={Zhao, Haoran and Pan, Fengxing and Ping, Huqiuyue and Zhou, Yaoming},
  journal={arXiv preprint arXiv:2311.15033},
  year={2023}
}

@article{tian2025uavs,
  title={{UAV}s meet {LLM}s: Overviews and perspectives towards agentic low-altitude mobility},
  author={Tian, Yonglin and Lin, Fei and Li, Yiduo and Zhang, Tengchao and Zhang, Qiyao and Fu, Xuan and Huang, Jun and Dai, Xingyuan and Wang, Yutong and Tian, Chunwei and others},
  journal={Information Fusion},
  volume={122},
  pages={103158},
  year={2025},
  publisher={Elsevier}
}

@inproceedings{brooks2023instructpix2pix,
  title={Instructpix2pix: Learning to follow image editing instructions},
  author={Brooks, Tim and Holynski, Aleksander and Efros, Alexei A},
  booktitle={Computer Vision and Pattern Recognition  },
  year={2023}
}

@inproceedings{zhou2024multimodal,
  title={Multimodal situational safety},
  author={Zhou, Kaiwen and Liu, Chengzhi and Zhao, Xuandong and Compalas, Anderson and Song, Dawn and Wang, Xin Eric},
  booktitle={International Conference on Learning Representations  },
  year={2025}
}

@article{he2024everyday,
  title={Everyday object meets vision-and-language navigation agent via backdoor},
  author={He, Keji and Chen, Kehan and Bai, Jiawang and Huang, Yan and Wu, Qi and Xia, Shu-Tao and Wang, Liang},
  journal={Advances in Neural Information Processing Systems},
  volume={37},
  pages={49684--49705},
  year={2024}
}

@inproceedings{zhan2025beat,
  title={BEAT: Visual Backdoor Attacks on VLM-based Embodied Agents via Contrastive Trigger Learning},
  author={Zhan, Qiusi and Ha, Hyeonjeong and Yang, Rui and Xu, Sirui and Chen, Hanyang and Gui, Liangyan and Wang, Yu-Xiong and Zhang, Huan and Ji, Heng and Kang, Daniel},
  booktitle={The Fourteenth International Conference on Learning Representations},
  year={2025}
}

@inproceedings{vqa2,
  title     = {Making the {V} in {VQA} Matter: Elevating the Role of Image Understanding in Visual Question Answering},
  author    = {Goyal, Yash and Khot, Tejas and Summers-Stay, Douglas and Batra, Dhruv and Parikh, Devi},
  booktitle = {Proceedings of the IEEE Conference on Computer Vision and Pattern Recognition},
  year      = {2017}
}

@article{vla2025survey,
  title={Vision-language-action models for robotics: A review towards real-world applications},
  author={Kawaharazuka, Kento and Oh, Jihoon and Yamada, Jun and Posner, Ingmar and Zhu, Yuke},
  journal={IEEE Access},
  year={2025},
  publisher={IEEE}
}
\clearpage

\end{document}